# Academic Research Output Derivatives: Structuring Futures and Options on Research Output Index

## Amarendra Sharma[1]

Version 1.0

May 26, 2025

## Abstract

This paper explores an innovative financial concept—Academic Research Output Futures (AROFs)—which aim to bring the tools of market-based finance into the realm of academic research. At the heart of this idea lies a cash-settled futures contract tethered to a composite Research Output Index (ROI), a metric designed to quantify the research productivity of universities and research institutions. By allowing investors to take positions—speculative or hedging—on the projected trajectory of academic performance, AROFs open the door to a fundamentally new funding model. For universities, this could mean access to capital markets without relying solely on grants, donations, or government subsidies. Accompanying these futures are Academic Research Output Options (AROOs), providing a layer of strategic nuance and risk control, much like options in traditional financial markets. The paper delves into the architecture of the ROI, the mechanics of the derivative instruments, regulatory and legal challenges, and the broader strategic consequences of such a framework. If realized, this model could represent a significant departure from conventional research funding approaches—one that aligns scholarly output with financial incentives in ways both promising and provocative.

## 1. Introduction

### 1.1 The Funding Challenge in Academic Research

Academic research serves as the engine behind much of society's progress—fueling advances in technology, breakthroughs in medicine, and the formulation of informed public policy (see for example, NAP, 2003; Galea & Bibbins-Domingo 2025; FP, 2022; Mirvis, 2009). Yet ironically, the financial framework that supports this intellectual enterprise remains fragile, fragmented, and in many ways antiquated. The majority of funding still flows from competitive public grants, charitable donations, or endowment allocations—sources that are not only inconsistent and politically contingent but also poorly aligned with the extended timelines and inherent uncertainty that define rigorous scholarly work (see for example, Geuna & Martin, 2003; Dasgupta & David, 1994). This precariousness is particularly stark in the case of early-stage or interdisciplinary research, which often falls through the cracks of narrowly defined funding channels. At the same time, universities are increasingly caught in a bind: governments, private donors, and the public alike now demand demonstrable returns on research investment—productivity metrics, societal impact, and innovation pipelines—placing pressure on institutions to quantify what was once valued intrinsically (see for example, Berkes et al, 2024; Bromham et al, 2016; Leahy & Barringer, 2020; Edwards & Roy,2017; Galindo-Rueda, 2013). Despite this shift toward performance-based evaluation, the financial sector still lacks a standardized, tradable instrument that can represent academic output as a tangible economic asset. This absence not only reflects a blind spot in market design but also echoes earlier calls for the expansion of risk-sharing instruments into undercapitalized, high-value domains like education and research (see for example, Shiller, 2009).

---

[1] Department of Economics, Binghamton University, Binghamton, NY, USA. Email: aksharma@binghamton.edu. The views and opinions expressed in this paper are solely those of the author and do not reflect the official policy or position of Binghamton University or any of its faculty or staff.

## 1.2 Contribution of this Paper

This paper introduces a new breed of financial derivative—Academic Research Output Futures (AROFs)—conceived as a market-oriented mechanism to both monetize and manage the risks associated with academic productivity. Structured as cash-settled contracts, each AROF is anchored to a multifaceted Research Output Index (ROI) that synthesizes various indicators of scholarly performance: peer-reviewed publications, field-normalized citation metrics, external grant funding, patent activity, and broader measures of societal impact. The conceptual and empirical groundwork for such indices can be traced through a diverse literature—from Bollen et al. (2009) and Lorden & Martin (2000) to more recent contributions by Keshavarz-Fathi et al. (2023) and Pal & Rees (2022), though critiques like Van Raan (2005) underscore the epistemic and methodological caveats still lurking beneath these composite metrics.

By allowing investors to take directional positions on the trajectory of an institution's research performance, AROFs catalyze the formation of a structured, liquid market around the latent potential of academic institutions. Much like traditional futures contracts in commodities or equities, these instruments provide a platform not only for speculation but also for hedging exposure to fluctuating research outcomes. The addition of Academic Research Output Options (AROOs) enhances this toolkit, enabling investors and universities alike to engage in strategic risk management and performance-aligned financing, drawing on the foundational insights of financial engineering à la Black & Scholes (1973) and Merton (1973).

The paper proceeds by laying out the theoretical framework for AROFs, detailing the construction logic behind the ROI, the settlement protocols for contracts, and the legal and institutional architecture necessary to sustain such a market. Beyond design and implementation, it proposes avenues for monetization and explores illustrative use cases across institutional types. In doing so, this work aims to bridge the traditionally siloed domains of academic policy, finance, and impact investing—offering a bold, market-based approach to solving one of the academia's most persistent dilemmas: how to fund research in a way that is both scalable and sustainable.

## 2. Theoretical Framework

The theoretical architecture underpinning AROFs rests at the intersection of financial economics, innovation theory, and the economics of public goods. At its conceptual core lies a deceptively simple but powerful proposition: academic research output constitutes a latent asset—one whose value materializes gradually and unevenly, often eluding traditional mechanisms of recognition and reward (see for example, Arrow, 1962; Stiglitz, 1999). By translating this intangible, forward-looking value into a tradable financial instrument, the AROF framework introduces a dual function—price discovery and risk allocation—within the largely opaque and undercapitalized research economy.

An institution's expected Research Output Index (ROI), serving as the pricing benchmark for AROFs, encapsulates its potential to generate outputs that carry academic merit, technological relevance, or societal significance. This framing resonates strongly with the logic of endogenous growth theory, wherein innovation and knowledge accumulation are not exogenous shocks but core drivers of sustained economic advancement (see for example, Romer, 1990). Just as equity markets discipline and finance corporate behavior, AROFs hold the promise of aligning incentives between research institutions and capital providers—creating a feedback loop in which performance, capital flow, and strategic orientation are mutually reinforcing.

Perhaps most critically, the AROF architecture remains sensitive to the public-good character of knowledge production. The ROI is constructed to reflect external signals of productivity—citations, funding, patents, or media visibility—without asserting ownership over the research itself. This design choice preserves the norms of open science and academic freedom (see for example, David, 2003), while embedding a layer of financial accountability that has long been absent from the scholarly ecosystem. In doing so, AROFs offer a rare equilibrium: a means of capitalizing research performance without commodifying the core epistemic function of the academy.

## 3. Construction of the Research Output Index (ROI)

The Research Output Index (ROI) serves as the quantitative backbone of the AROF framework, offering a multidimensional portrait of an academic institution's productivity over a specified time horizon. Rather than relying on a single proxy such as publication count or grant volume, the ROI is crafted as a composite metric that integrates five distinct, yet interrelated, dimensions of scholarly output:

1. Publication Output (P) – The raw volume of peer-reviewed journal articles, discipline-normalized and impact-factor weighted to correct for disparities across fields.
2. Citation Impact (C) – A field-weighted citation metric computed over a rolling five-year period, capturing influence rather than just activity.
3. Grant Income (G) – Aggregate competitive research funding, scaled to institutional size to account for economies of scope and scale.
4. Innovation Output (I) – Patent filings, license agreements, and the emergence of spin-off ventures, reflecting knowledge transfer and commercialization capacity.
5. Societal Impact (S) – Indicators of real-world resonance, including citations in policy documents, media visibility, and metrics of public engagement.

These five components are blended into a single index using the formula:

$$ROI = w_1 P + w_2 C + w_3 G + w_4 I + w_5 S$$

The weighting coefficients ($w_1$ through $w_5$) are not arbitrarily assigned but are determined through a deliberative process managed by an independent standards board comprising experts from academia, capital markets, and policy domains. Data is submitted annually by institutions using a standardized reporting framework, audited independently, and benchmarked against discipline-specific norms to ensure both comparability and credibility. In many ways, this approach echoes existing evaluative methodologies like Scimago, Star Metrics, and other global ranking systems, but with a critical difference: here, the ROI is not an end in itself—it's an economic signal embedded in a tradeable financial asset, one that could reshape how research is financed and valued.

## 4. Derivative Structure of AROFs and AROOs

Academic Research Output Futures (AROFs) are structured as standardized, cash-settled futures contracts linked directly to the Research Output Index (ROI) of a participating university or research institution. Each contract specifies a fixed notional value—for instance, $1,000 per ROI point—and is traded on a regulated electronic exchange. Settlement occurs annually, contingent on the final, independently verified ROI figure. The mechanism is straightforward but powerful in its implications for both speculation and risk management.

To illustrate: imagine an investor acquires 50 AROF contracts on University X, whose current ROI stands at 100. If, at the end of the year, the ROI rises to 115, the net gain realized by the investor is:

$$(115 - 100) \times 50 \times \$1,000 = \$750,000.$$

On the other hand, if performance slips and the ROI declines to 95, the investor faces a downside

exposure of:

$$(95 - 100) \times 50 \times \$1,000 = -\$250,000.$$

Complementing the futures are Academic Research Output Options (AROOs)—European-style derivatives that enable more nuanced strategies for managing academic performance risk. Institutions or investors may use calls to benefit from unexpected surges in research output or puts to guard against performance deterioration. The pricing of these options adheres to the classic frameworks pioneered by Black & Scholes (1973) and Merton (1971), though the models are recalibrated to reflect the unique volatility dynamics of the ROI.

**Pricing AROOs: Theory, Intuition, and a Worked Example**

Academic Research Output Options (AROOs) are forward-looking instruments, and like traditional equity options, their value hinges on expectations about future performance—except here, the performance is not of a stock, but of a university's or lab's research productivity, captured by a Return on Research Investment index (ROI).

An adapted version of the Black-Scholes model can be used to price them with one important substitution: instead of modeling stock price behavior, we treat the ROI index as the underlying asset.

The model tells us the fair price of a call option on this ROI index using the following formula:

$$C = S_0 N(d_1) - K e^{-rT} N(d_2)$$

Where the terms $d_1$ and $d_2$ are represented by the following expressions:

$$d_1 = \frac{ln\left(\frac{S_0}{K}\right) + \left(r + \frac{\sigma^2}{2}\right) T}{\sigma \sqrt{T}}$$

$$d_2 = d_1 - \sigma\sqrt{T}$$

Where:

$S_0$: Current level of the ROI index (e.g., 100)

$K$: Strike ROI—like a research output goalpost

$T$: Time to maturity (years)

$r$: Risk-free annualized interest rate (say, from Treasury bills)

$\sigma$: Estimated volatility of the ROI index (based on past fluctuations in metrics like citations, patents, grant inflows)

$N(d_1)$ & $N(d_2)$: The cumulative distribution functions for standard normal variables

Analogously, the formula for a put option is:

$$P = Ke^{-rT}N(-d_2) - S_0N(-d_1),$$

where $P$ is the price of the put option, and other variables remain the same.

To illustrate how these AROO's work, let us suppose that XYZ Lab has an ROI index currently at 100. An investor is bullish on the lab's trajectory—perhaps because XYZ just launched a $1 billion AI initiative or secured major DARPA funding. The investor buys a 3-year European-style call option with a strike of 110.

**Assumptions:**

$S_0 = 100$, $K = 110$, $T = 3$, $r = 0.03$ (3% annual risk-free rate), $\sigma = 0.18$ (18% annual volatility based on ROI data trends)

First, compute $d_1$ and $d_2$:

$$d_1 = \frac{\ln\left(\frac{100}{110}\right) + (0.03 + 0.5 * 0.18^2) * 3}{0.18 * \sqrt{3}} = \frac{-0.0953 + 0.1386}{0.3118} \approx 0.1389$$

$$d_2 = 0.1389 - 0.3118 \approx -0.1729$$

Using a standard normal table:

$$N(d_1)) \approx 0.5552$$

$$N(d_2)) \approx 0.4315$$

Now plug into the Black-Scholes formula:

$$C = 100 * 0.5552 - 110 * e^{-0.09} * 0.4315 = 55.52 - 43.37 \approx 12.15$$

Thus, the fair price of the call option is $12.15 per contract.

Analogously, the fair price of the put option using same assumptions can be calculated as $12.68 per contract.[2]

**Payoff Scenarios at Expiry**

Let's say the investor buys 1,000 contracts and pays a total premium of $12,150.

If ROI index rises to 125 at maturity:

$$profit = (125 - 110) * 1000 - 12150 = 15000 - 12150 = \$2850$$

If ROI index ends at 108:

Option expires worthless. Loss = Premium = $12,150.

---

[2] The Python code to calculate Call and Put Option prices using the above formula is provided in the appendix. The user can change the values of variables used in the assumptions in the above example in the code to recalculate a different scenario.

**Interpretation**

AROOs behave like regular call options: limited downside (the premium) and unlimited upside (bounded by research success). But unlike tech stocks or commodities, the ROI index is powered by peer-reviewed publications, intellectual property, research citations, and grant flows.

Estimating volatility ($\sigma$) might involve historical time series of NSF funding trends, publication databases (like Scopus or Web of Science), or machine learning models of research impact trajectory.

AROFs and AROOs together form a sophisticated toolkit for transforming raw research performance into actionable financial instruments—blending capital markets precision with the uncertain, long-horizon nature of knowledge production.

## 5. Monetization Pathways for Universities

Universities stand to gain immensely from Academic Research Output Futures (AROFs), tapping into a multifaceted array of financial strategies that bolster capital inflows, mitigate risk, and uphold the sanctity of scholarly pursuits.

### 5.1 Exchange-Traded Issuance

Universities may tap directly into capital markets by issuing Academic Research Output Futures (AROFs) via licensed derivative exchanges, converting latent academic potential into immediate, flexible liquidity. The revenue from these initial offerings provides a fresh infusion of capital— non-dilutive and performance-contingent—that can be channeled into ambitious research agendas, laboratory expansions, or cross-disciplinary ventures. Consider, for instance, a prominent STEM institution launching 1,000 AROF contracts at a benchmark ROI of 100. If the market perceives catalytic developments on the horizon—such as the launch of a federally funded AI research center or a strategic alliance with the private sector—investors may bid up the contract price in anticipation, injecting millions into the university's operational pipeline before a single paper is published.

This mechanism particularly favors institutions with strong reputational equity or credible signals of forthcoming impact. Transparent disclosures, third-party audits, and forward guidance akin to corporate earnings calls can help academic issuers shape investor expectations and strengthen market confidence.

## 5.2 Hedging Tools

Volatility is a persistent feature of the academic funding landscape, especially for universities dependent on state appropriations or donor-driven endowments. Academic Research Output Options (AROOs)—structured as options on the ROI—provide a financial buffer against such unpredictability. A public university bracing for legislative budget cuts, for example, might acquire put options to insure against a potential decline in its ROI. Should the index fall from 100 to 90, the payout from the option can soften the blow, plugging fiscal gaps and maintaining research continuity.

Beyond hedging, AROOs offer intertemporal budgeting tools for research offices and endowment managers. Institutions anticipating a surge in research visibility—perhaps driven by a high-impact publication series or the announcement of major grant awards—could purchase call options on their own ROI, monetizing future success and using the proceeds to fund long-term initiatives. In essence, AROOs turn expectations of excellence into tangible financial levers, aligning institutional strategy with market-based foresight.

**5.3 Private Placement and Custom Structures**

Not all AROF contracts must be funneled through public markets; tailor-made arrangements with strategic investors—including alumni networks, university-backed venture vehicles, or mission-aligned philanthropic foundations—can be structured through private placements. These off-exchange deals allow for customization in maturity, performance metrics, and payoff structures, thereby accommodating niche research goals or specialized funding constraints.

Imagine a top-tier engineering school partnering with a climate-focused foundation on a five-year AROF series tied to innovation in renewable energy. The performance metrics might emphasize patent disclosures, publication impact in top-tier green technology journals, or licensing deals with cleantech startups. In doing so, the institution and its benefactors co-construct a financial instrument that translates shared values into co-aligned incentives.

AROFs and their derivative extensions offer universities a potent and flexible financial toolkit—one that not only diversifies revenue streams but also embeds performance accountability and market validation into the fabric of academic funding. By unlocking the embedded economic value of research productivity, institutions can reimagine how excellence is both pursued and financed.

**6. Strategic Use Cases**

To showcase the tangible benefits and risk-reward dynamics of AROFs and AROOs, let's dive into three vivid, hypothetical scenarios.

**6.1 Use Case 1: Capitalizing on Research Momentum**

Imagine a mid-sized university, long respected but rarely in the spotlight, that begins to carve a niche in the competitive field of neuroscience. Following a strategic recruitment wave that brings in marquee faculty and a multimillion-dollar investment in laboratory infrastructure, institutional leaders anticipate a significant uptick in research output—both in volume and influence. Rather than wait for traditional grant cycles to catch up, the university issues Academic Research Output Futures (AROFs), pegged to its current ROI baseline of 95, raising $10 million in near-term, non-dilutive funding. These funds are immediately deployed to complete a state-of-the-art brain imaging facility—cementing the university's capacity to lead in a frontier discipline.

As projected, over the subsequent three years, the institution's ROI rises steadily, reaching 120 on the back of prolific publications, increased citations, and growing media visibility. Investors realize handsome returns, while the university reaps reputational dividends that attract further capital and talent. This feedback loop—where institutional foresight meets financial engineering—demonstrates how AROFs can convert academic momentum into scalable, self-reinforcing value.

## 6.2 Use Case 2: Smoothing Research Budgets

In a very different context, consider a sprawling public university situated in a politically volatile state where electoral cycles routinely reshape budget priorities. Anticipating a downturn in appropriations tied to shifts in government leadership, university administrators turn to Academic Research Output Options (AROOs) as a preemptive financial hedge. They purchase put options with a strike level set at an ROI of 100, effectively insuring against a decline in measurable research output.

When post-election budget cuts materialize and the university's ROI slides to 92, the AROO positions trigger payouts that help offset the immediate fiscal shortfall. The inflow preserves continuity in research programs, prevents layoffs, and maintains critical infrastructure through the turbulence. In this scenario, derivatives act not as speculative tools, but as stabilizers—financial shock absorbers capable of smoothing budgetary cycles and safeguarding academic resilience in environments where uncertainty is the only constant.

## 6.3 Use Case 3: Sustaining Humanities and Liberal Arts

Departments in the humanities and liberal arts, often underfunded compared to STEM fields, can benefit from AROF markets by leveraging non-traditional impact metrics such as public scholarship reach, book citations, cultural institution collaborations, and policy influence. A liberal arts college may issue AROFs benchmarked against an ROI index that includes metrics like humanities journal impact scores, op-ed publication rates, and public lecture attendance.

While these outputs are less likely to result in patents or commercial spin-offs, they do unlock monetizable pathways such as philanthropic endowments, cultural grants, and executive education programs in ethics, communication, and leadership. For instance, increased public engagement and reputational prestige can enhance appeal to major donors or attract funding from humanities-focused foundations. High-visibility public scholarship and policy contributions may also lead to paid consulting roles or content partnerships with media, education, and civic organizations.

Through this indirect but credible feedback mechanism, universities can translate academic visibility into fundable value, making humanities AROFs a feasible and innovative tool for sustaining liberal education.

## 7. Risk Management and Payout Strategies for Universities

The viability of AROF markets hinges critically on the ability of academic institutions to honor their obligations under high ROI conditions. When the ROI index exceeds the agreed-upon strike level, universities as AROF issuers are liable for payout differentials to investors. This exposure necessitates robust financial planning and institutional safeguards to maintain market integrity and institutional solvency.

## 7.1 Monetized Research Returns

One sustainable approach lies in the monetization of research outputs—universities increasingly derive revenue from patent licensing, start-up equity participation, and industrial R&D contracts (see for example, Mowery et al., 2015). For instance, a medical school's licensing of a breakthrough diagnostic technology to a pharmaceutical firm could yield multi-million-dollar royalties. These earnings, inherently correlated with spikes in ROI, serve as endogenous hedges that naturally fund derivative payouts.

## 7.2 Reserve Funds and Hedging Instruments

A prudent institution may allocate a fraction of its upfront AROF proceeds to a dedicated reserve fund, earmarked for future settlement obligations. Alternatively, universities may hedge their exposure by purchasing AROO puts—effectively acquiring insurance against adverse movements in their own ROI—or by participating in pooled risk consortia and third-party reinsurance arrangements. These mechanisms distribute and mitigate institutional risk across broader financial ecosystems.

### 7.3 AROF Issuance Revenue as Liability Buffer

Capital raised via AROF issuance is itself a strategic asset. By investing proceeds in low-risk, liquid instruments—e.g., Treasury bonds or endowment-grade vehicles—universities generate steady returns that can be partially sequestered for liability coverage. This model parallels the insurance sector's use of premium income to underwrite future claims (see for example, Rejda & McNamara, 2017).

### 7.4 Structural Risk Controls in Contract Design

Well-structured AROFs may incorporate caps, collars, and trigger thresholds to limit university liabilities. A contract may, for example, cap payouts at 120% of strike ROI or impose aggregate annual liability ceilings. Such guardrails introduce predictability to financial exposure, facilitating budgetary planning and enhancing institutional confidence in engaging with the market (see for example, Hull & Basu, 2016)

### 7.5 Third-Party Underwriting

Universities may also shift some risk to financial intermediaries. Investment banks, reinsurance firms, or mission-aligned hedge funds could underwrite AROF tranches in exchange for premiums or structured participation rights. This model draws inspiration from bond insurance and credit default swaps, extending credibility and liquidity to the nascent academic derivatives ecosystem (see for example, Hull & Basu, 2016).

In aggregate, these instruments and strategies furnish universities with a sophisticated risk toolkit—enabling participation in AROF markets without jeopardizing fiscal sustainability or academic independence.

## 8. Governance and Exchange Infrastructure

The advent of AROFs and AROOs ushers in a tangle of legal and compliance conundrums, driven by the peculiar nature of their underlying assets—academic outputs that shine with intellectual promise but yield economic value only indirectly or over time.

### 8.1 Securities Classification

At the heart of any financial innovation lies a fundamental legal inquiry: how will it be classified under prevailing securities law? In the case of Academic Research Output Futures (AROFs), their cash-settled architecture and dependence on a quantifiable research index strongly suggest they may be construed as security-based swaps or futures contracts under U.S. statutes, particularly the Securities Exchange Act of 1934. This classification would almost certainly invoke the jurisdiction of the Securities and Exchange Commission (SEC) and possibly the Commodity Futures Trading Commission (CFTC) under Dodd-Frank Act provisions. To preempt compliance pitfalls, legal architects must coordinate early with regulatory bodies, shaping the product in a way that ensures both investor protection and administrative feasibility—without inadvertently triggering undue regulatory frictions.

### 8.2 Exchange Infrastructure

For AROFs to gain traction beyond conceptual intrigue, they must be embedded within the machinery of institutional finance—namely, regulated derivative exchanges. Listing on established platforms such as the Chicago Mercantile Exchange (CME) or the Intercontinental Exchange (ICE) would confer credibility, enforce standardized terms, and facilitate real-time clearing and trade transparency. A dedicated index—such as the Academic ROI—would act as the referential anchor for settlement, its construction overseen by academically credible and financially rigorous entities. Data veracity, dissemination protocols, and regulatory audit trails would all hinge on a transparent, well-governed exchange infrastructure that bridges academia and capital markets.

## 8.3 Data and Privacy Standards

At the operational core of AROFs lies a data-intensive process: the construction of robust, multi-dimensional Research Output Indices (ROIs). To preserve market integrity and institutional trust, data inputs—ranging from publication metrics and normalized citations to grant flows and technology transfers—must be auditable, standardized, and privacy-compliant. Given the sensitivity of these datasets, institutions must implement data governance protocols that ensure anonymization, secure transmission, and consensual aggregation. The likely emergence of third-party data custodians—possibly in the form of nonprofit consortia or university-backed entities—would serve to centralize data management and guarantee methodological consistency.

## 8.4 Institutional Governance

For universities, the decision to engage with AROFs entails more than just financial engineering—it requires the scaffolding of a new internal governance paradigm. To mitigate reputational risk and prevent moral hazard, institutions must erect firewalls between market incentives and academic integrity. Annual ROI disclosures should be vetted by independent audit panels, and AROF performance must remain explicitly decoupled from faculty evaluations, promotion criteria, or researcher hiring. A deliberate separation of speculative market outcomes from day-to-day academic operations is essential, and ethical oversight—possibly under an institutional research board—must be embedded into every phase of participation.

## 8.5 International Variance

The cross-border ambitions of AROFs necessitate early engagement with the fragmented global regulatory landscape. Jurisdictions such as the European Union, United Kingdom, Singapore, and Canada may interpret performance-based academic instruments differently, depending on how local statutes define securities, public goods, and data sovereignty. As such, a roadmap for regulatory harmonization—likely through multilateral memoranda of understanding and sandbox frameworks—will be vital. Without such alignment, AROFs may face prohibitively uneven compliance thresholds, undermining their utility in global capital markets.

## 9. Legal and Regulatory Framework

### 9.1 Jurisdictional Landscapes and Regulatory Classification

The legal architecture governing Academic Research Output Futures (AROFs) and Academic Research Output Options (AROOs) is inherently complex, straddling the boundaries of financial innovation, public policy, and global capital regulation. Positioned at the confluence of non-traditional asset classes, event-contingent derivatives, and index-linked financial instruments, these contracts challenge conventional regulatory typologies. Their classification will determine not only the scope of permissible market activity but also the attendant obligations around disclosure, capital adequacy, and investor eligibility.

### United States (U.S.)

In the U.S., regulatory jurisdiction would almost certainly fall to the Commodity Futures Trading Commission (CFTC), given the cash-settled structure and the use of a non-financial benchmark—the Research Output Index (ROI)—as the underlying. Under the Commodity Exchange Act (CEA), several considerations would influence classification:

**Nature of the Index**: If the ROI aggregates data across departments or institutions, it could qualify as a broad-based index, rendering AROFs functionally similar to equity index futures.

**Character of the Underlying**: Given that ROI reflects intangible academic performance rather than a physical or financial asset, AROFs may be analogized to event-based derivatives, akin to contracts on electoral outcomes or environmental metrics.

**Settlement Mechanism**: As fully cash-settled instruments without physical delivery obligations, AROFs may be structured under Part 36 Exempt Markets or registered for public trading on Designated Contract Markets (DCMs) if sufficiently standardized.

For AROOs—structured as options on ROI values—regulation under Title VII of the Dodd-Frank Act would likely apply, especially for institutional use. Early pilots may benefit from no-action relief or exemptive guidance, particularly if structured for limited, experimental deployment. If retail access is contemplated, Securities and Exchange Commission (SEC) scrutiny may be triggered under the Howey Test, especially if the offering resembles an investment contract with profit expectations derived from institutional performance.

**European Union (EU)**

Within the EU, oversight would be administered by the European Securities and Markets Authority (ESMA), primarily through the Markets in Financial Instruments Directive II (MiFID II) and the European Market Infrastructure Regulation (EMIR). AROFs and AROOs would likely be classified as:

Contracts for Difference (CFDs) or index-based derivatives, requiring registration and trading via Multilateral Trading Facilities (MTFs) or Organized Trading Facilities (OTFs).

Instruments subject to EMIR obligations, including central clearing, transaction reporting, and risk mitigation measures such as bilateral margining for non-centrally cleared trades.

Academic institutions may qualify as non-financial counterparties (NFCs) under EMIR, depending on the volume and nature of their engagement. Any cross-border issuance would need to comply with passporting rules or be granted equivalency recognition for third-country platforms.

## United Kingdom (UK)

Following Brexit, regulatory stewardship in the UK has shifted to the Financial Conduct Authority (FCA). Although mirroring EU frameworks in many respects, the UK approach may differ in implementation. The FCA retains product intervention powers, particularly in scenarios where retail market participation is allowed. Clearinghouses and exchanges would be subject to UK EMIR prudential standards, including margin, reporting, and operational resilience requirements. The regulatory sandbox program offers a tailored pathway for early stage testing of AROFs/AROOs in a controlled legal environment.

## Other Jurisdictions

**Canada**: The Canadian Securities Administrators (CSA) and provincial regulators would oversee AROFs depending on whether they are offered as exchange-traded or OTC instruments. Treatment could vary significantly by province.

**India**: The Securities and Exchange Board of India (SEBI) enforces stringent rules on derivatives tied to non-financial indices. A co-regulatory or sandbox model in collaboration with public research institutions may be the only viable route forward.

**Singapore & Australia**: Regulators such as Monetary Authority of Singapore (MAS) and Australian Securities and Investments Commission (ASIC) are generally conducive to fintech experimentation. Both maintain well-regarded sandbox frameworks that could facilitate early-stage issuance with adaptive licensing.

## 9.2 Recommendations for Legal Launch Strategy

Given the novelty and hybrid nature of AROFs, a staged, jurisdiction-sensitive deployment is advisable. Key steps include:

**Sandbox Piloting**: Launch within regulatory sandboxes in innovation-forward jurisdictions (e.g., FCA, MAS) to allow iterative refinement of legal structures and compliance protocols.

**Pre-Launch Legal Opinions**: Secure authoritative legal interpretations from financial law specialists in the U.S., EU, UK, and Asia-Pacific regions to clarify classification and anticipate regulatory obligations.

**Tiered Market Access**: Restrict initial offerings to Qualified Institutional Buyers (QIBs) or accredited investors to mitigate early compliance risk while gathering performance data.

**Structured Exemptions**: Work proactively with regulators to develop public-benefit carveouts or customized exemptions, drawing on precedents from environmental derivatives, catastrophe bonds, or climate risk instruments.

In summary, the successful operationalization of AROFs and AROOs will hinge on regulatory ingenuity, cross-border coordination, and the development of tailor-made legal frameworks that recognize the unique intersection of academic productivity and financial engineering. With thoughtful structuring, these instruments can evolve from speculative novelty to a globally trusted mechanism for research finance.

## 10. Potential Challenges and Mitigation Strategies

While AROFs offer a groundbreaking mechanism for channeling capital into academic research, their implementation is not without formidable obstacles. These span a gamut of practical, ethical, and technical dimensions, ranging from institutional behavior and data integrity to broader social concerns about the commodification of knowledge. Recognizing and addressing these complexities is critical to building a resilient and credible academic derivatives market.

### 10.1 Risk of Perverse Incentives and Academic Gaming

The introduction of financial instruments tied to academic output creates potential for distortionary incentives. By monetizing research productivity, institutions may feel compelled to prioritize metrics amenable to rapid gains—such as publication volume or grant acquisition—at the expense of depth, originality, or long-horizon inquiry. This may entrench metric-driven behaviors that undermine the epistemic foundations of academia (see for example, Haack 2022; Mathies et al., 2020).

**Mitigation Strategy:** The ROI must incorporate multidimensional, time-weighted, and field-normalized metrics that reward not just volume but verifiable impact, reproducibility, and scholarly rigor. Inclusion of peer-reviewed quality indicators, third-party evaluations, and safeguards against disciplinary bias is crucial. An independent governance board should oversee regular audits and dynamic recalibration of the index to preserve its integrity over time.

## 10.2 Asymmetric Information and Insider Trading Risk

Academic institutions inherently possess privileged knowledge regarding their own pipelines—upcoming publications, major grant submissions, or faculty recruitment—all of which may materially affect the ROI. This asymmetry mirrors insider trading risks in conventional securities markets (see for example, Lakonishok & Lee, 2001) and poses a threat to the fairness and transparency of AROF trading.

**Mitigation Strategy:** Robust governance protocols should mirror the SEC's insider trading rules, including disclosure obligations, pre-clearance of trades, and blackout periods prior to index updates. Surveillance systems must be equipped to detect anomalous trading patterns, while third-party audits can ensure that the release of sensitive research data is appropriately timed and equitably disclosed.

## 10.3 Liquidity and Market Adoption

No matter how elegantly designed, a market without sufficient liquidity will falter. Illiquidity can manifest as volatile pricing, thin order books, and inefficient price discovery, especially in the early phases of market deployment (see for example, Wadesango et al, 2017). This is particularly salient in academic markets, where investor familiarity and confidence may be initially limited.

**Mitigation Strategy:** Early-stage liquidity should be cultivated through strategic partnerships with research-intensive universities, mission-aligned investors such as ESG funds, and major endowments. Designating market makers and offering participation incentives during the pilot phase can help stabilize spreads and ensure smoother price formation. A centralized exchange architecture with transparent protocols may further reduce friction.

## 10.4 Data Collection and Verification

AROFs are only as credible as the data underlying them. Variability in reporting standards, inconsistent data streams, or time lags across academic institutions could impair the construction and credibility of the ROI (see for example, Florida Board of Governors 2024; SensisData n.d.). Such data anomalies could render contract settlement contentious or erode investor trust.

**Mitigation Strategy:** A unified data infrastructure—featuring API-integrated feeds from bibliometric and grant-tracking databases such as Scopus, NIH RePORTER, and Web of Science—is essential. Field normalization and timestamping protocols should be enforced, while a distributed auditing framework involving certified third parties or peer institutions can enhance data verifiability and robustness.

## 10.5 Legal and Reputational Risk

Universities may balk at exposing their academic trajectories to public speculation, particularly if falling ROI scores are misconstrued as a signal of institutional decline. This could affect faculty morale, applicant perceptions, donor confidence, or even accreditation outcomes (see for example Deloitte 2023; Xapien n.d. for risks facing universities).

**Mitigation Strategy:** Institutional participation should remain opt-in, with clear guidelines on usage and the ability to exit without penalty. Contracts may initially be restricted to private placements or internal hedging mechanisms to insulate public perception. Transparency around the scope, methodology, and interpretative limits of the ROI is critical for managing reputational risk.

**10.6 Ethical Concerns and Public Perception**

The transformation of scholarly output into a tradable commodity inevitably raises ethical questions. Critics may argue that such financialization devalues the intrinsic motivations that underpin academic inquiry and encourages transactional over collaborative or curiosity-driven behaviors (see for example, Caulfield & Ogbogu 2015; Mensah, n.d.).

**Mitigation Strategy:** Framing is paramount. AROFs and AROOs should be positioned as tools for augmenting research sustainability and public benefit, not merely speculative gain. Proceeds from contracts should be earmarked for reinvestment into foundational or underserved research areas. Continuous dialogue with ethics boards, faculty senates, and student organizations can help ensure that implementation aligns with academic values and community expectations.

**10.7 Issuance and Market Infrastructure**

The question of who should issue AROFs and AROOs is not merely logistical—it is foundational to the credibility, functionality, and regulatory viability of the entire market. Unlike traditional securities issued by corporations or commodities managed by producers, AROFs require a neutral, trustworthy, and academically aligned issuer. In this context, issuance should be the purview of a dedicated academic derivatives exchange or a licensed financial intermediary operating under the supervision of regulatory authorities such as the CFTC or FCA. These entities would standardize contract terms, manage settlements, and ensure compliance with index governance protocols.

Universities would not issue contracts themselves, both to avoid perceived conflicts of interest and to maintain institutional independence from speculative financial activities. Instead, their role would center on data provision and voluntary participation in the ROI framework. Analogous to the structure of event-based or weather derivatives, the underlying index—curated by a consortium of bibliometric analysts, economists, and domain experts—would serve as the reference metric. Contracts could then be listed and traded by institutional investors, philanthropic foundations, or ESG-aligned funds seeking exposure to research productivity as an asset class (see for example, Benth, Benth, and Koekebakker 2008 for futures market in electricity).

**Mitigation Strategy:** Early issuance should be limited to sandboxed exchanges within innovation-friendly jurisdictions to allow for regulatory flexibility. Legal opinions should be obtained to classify AROFs appropriately under existing derivatives law, and initial offerings should target accredited investors to manage systemic risk. The creation of a not-for-profit governing body, akin to the role of the CME Group in agricultural futures, could help institutionalize trust, manage disclosures, and support the broader mission of sustainable research financing.

## 11. Conclusion: Rethinking Research Capital Through Financial Innovation

Academic Research Output Futures (AROFs) mark a radical reimagining of how society might channel capital toward the engine of knowledge production. By operationalizing a transparent, quantifiable, and tradeable measure of academic productivity—then embedding it within a legally robust derivative framework—AROFs stand to resolve enduring structural inefficiencies in how research is financed. No longer must universities depend solely on erratic appropriations or philanthropic largesse; instead, they can access scalable, risk-adjusted capital markets that reward long-term scholarly output.

For investors, AROFs offer more than speculative utility—they present an opportunity to align financial return with intellectual and societal impact, enabling a new class of mission-driven, data-anchored financial instruments. Realizing this vision, however, will demand more than ingenuity.

It calls for coordinated institutional governance, uncompromising data standards, and regulatory architectures that are both vigilant and adaptive.

Yet, if executed with foresight and integrity, the rewards could be profound. In anchoring the production of knowledge to the machinery of capital markets, AROFs chart a future in which research becomes as liquid, incentivized, and dynamically priced as any core asset class— a future where investment in ideas is not merely idealistic, but infrastructural.

**Conflict of Interest Statement:** The author maintains an investment portfolio, which includes trading in publicly listed equities and options. The author affirms that there are no financial or personal relationships that could be perceived as influencing the research presented in this paper.

# Appendix

# Python Code to Calculate Black-Scholes Formula for Call and Put Options

```python
import math
from scipy.stats import norm

def black_scholes(S, K, T, r, sigma, option_type="call"):
    """
    Calculate the Black-Scholes option price for a European call or put option.

    Parameters:
    S (float): Current stock price
    K (float): Strike price
    T (float): Time to expiration (in years)
    r (float): Risk-free interest rate (annualized)
    sigma (float): Volatility of the stock (annualized)
    option_type (str): "call" for call option, "put" for put option

    Returns:
    float: Option price
    """
    # Input validation
    if S <= 0 or K <= 0 or T <= 0 or sigma <= 0:
        raise ValueError("S, K, T, and sigma must be positive")
    if option_type.lower() not in ["call", "put"]:
        raise ValueError("option_type must be 'call' or 'put'")

    # Calculate d1 and d2
    d1 = (math.log(S / K) + (r + 0.5 * sigma ** 2) * T) / (sigma * math.sqrt(T))
    d2 = d1 - sigma * math.sqrt(T)

    # Debug prints
    print(f"d1: {d1:.4f}, d2: {d2:.4f}")
    print(f"norm.cdf(d1): {norm.cdf(d1):.4f}, norm.cdf(d2): {norm.cdf(d2):.4f}")

    # Calculate the option price based on the type
    if option_type.lower() == "call":
        price = S * norm.cdf(d1) - K * math.exp(-r * T) * norm.cdf(d2)
    else:  # put
        price = K * math.exp(-r * T) * norm.cdf(-d2) - S * norm.cdf(-d1)

    return round(price, 4)

def main():
    print("Starting main function")
    # Example parameters (as of May 25, 2025)
    S = 100      # Current stock price
    K = 110      # Strike price
    T = 3.0      # Time to expiration (3 years)
    r = 0.03     # Risk-free rate (3%)
    sigma = 0.18  # Volatility (18%)

    # Calculate call and put option prices
    call_price = black_scholes(S, K, T, r, sigma, "call")
    put_price = black_scholes(S, K, T, r, sigma, "put")

    # Print results
    print(f"Black-Scholes Option Prices:")
    print(f"Call Option Price: ${call_price}")
    print(f"Put Option Price: ${put_price}")

if __name__ == "__main__":
    main()
```